# “X-ray Coulomb Counting” to understand electrochemical systems

Chuntian Cao[1], Hans-Georg Steinrück[2,3,*]

[1]Artificial Intelligence Department, Brookhaven National Laboratory, Upton, NY 11973, USA

[2]Institute for a sustainable Hydrogen Economy (IHE-1), Forschungszentrum Jülich GmbH, An der Deutschen Welle 7a, 52428 Jülich, Germany

[3]Institute of Physical Chemistry, RWTH Aachen University, Landoltweg 2, 52074 Aachen, Germany

*: corresponding author: h.steinrueck@fz-juelich.de

Abstract: Electrochemical systems are important for a sustainable and defossilized energy system of the future. While accurate and precise, the corresponding electrochemical measurements, in which many reactions may occur simultaneously, often do not contain enough information to understand the underlying mechanism and processes. This information, however, is crucial towards rational materials and devices as well as process development and for inventing innovative concepts. In this perspective, we introduce explicitly the concept of “X-ray Coulomb Counting” in which X-ray methods are used to quantify on an absolute scale how much charge is transferred into which reactions during the electrochemical measurements. This allows to interpret the electrochemical measurements in detail and obtain the desired phenomenological and mechanistic understanding. We show a few recent examples from the Li-ion battery literature in which the concept of X-ray Coulomb Counting was employed to obtain foundational understanding.

Keywords: Electrochemistry, Li-ion batteries, X-ray absorption, X-ray Coulomb Counting, X-ray diffraction, X-ray reflectivity

## I. Introduction

Electrochemical systems are at the core of the transition to a sustainable and defossilized energy system of the future [1, 2]. The corresponding electrochemical devices include electrolyzers [3], batteries [4], supercapacitors [4], fuel cells [4], electrochemical water desalination reactors [5], and electrosynthesis reactors [6]. A typical electrochemical cell, the basis for all the referenced devices, is made of three electrodes, i.e., the working electrode (WE), the counter electrode (CE), and the reference electrode (RE), which are all immersed in an electrolyte [7]. These can be arranged in a variety of ways and a generic setup for a three-electrode cell is shown in Figure 1, which also illustrates a potentiostat, which is typically used in research to control the electrochemical cell. The voltage $E$ is controlled/measured between the WE and RE (or CE for a two-electrode cell); the current $I$ flows between the WE and CE, and is sometimes given per area, weight, or volume of the WE, yielding the current density $J$.

While the performance of the respective electrochemical device, which often does not contain a RE, depends on the interplay between all component (including cell design, reactor design, etc.), on a fundamental level one is typically interested in the electrochemical redox reactions occurring at or near the WE and the corresponding concentration changes of reactants. These reactions may include intercalation reactions, metal plating reactions, polymerization reactions, and alloying reactions, as well as solution species redox reactions; several reactions can in principle occur simultaneously. One may conceptualize one’s interest in the WE into the following two questions:

Q1. Which reaction(s) occur(s) at the WE?

Q2. How much of this/each reaction occurs?

One strives to answer these questions as a function of the external stimulus, which is typically current or voltage, but could also be time, temperature, etc. The answer to Q2 should ideally be quantitative on an absolute scale: how much charge / how many electrons are transferred (in Coulombs (C) or C/area or C/mass)? Depending on the type of reaction, one may add a third question, which may provide insight into “why” the reactions are occurring and the respective mechanisms:

Q3. During the reaction(s), what happens to the WE itself?

The purpose of this perspective is to rationalize why X-ray methods are particularly useful to answer these three questions with relatively high accuracy and precision. It is structured as follows: First, we briefly describe the conventional approach using electroanalytical methods. Then, we introduce explicitly and explain in detail the concept of "X-ray Coulomb Counting", i.e., the concept of how X-ray methods can be used to determine quantitatively on an absolute scale how much charge was transferred into which reaction during an electrochemical measurement. This is followed by a few recent examples, focusing somewhat on our own battery-related work. Finally, we provide a conclusion and an outlook.

## II. Electroanalytical methods

The archetypical electroanalytical method to investigate electrochemical devices is cyclic voltammetry (CV) [7-9]. CV is a potentiodynamic method, in which the voltage of the WE is varied between two vertex potentials; typically, the modulus of the time-derivative of the voltage, i.e., scan speed $v$, is kept constant. The corresponding current response is recorded and then typically plotted as a function of the voltage of the WE. A typical CV is shown in Figure 2. This CV was computed using Butler-Vollmer kinetics (code from [10] for macroelectrode) for a simple one-electron solution species redox reaction ($X^+ + e^- \rightleftharpoons X$, with formal potential $E_F^0$) and exhibits the typical "duck-shape" with a cathodic (reduction) peak and an anodic (oxidation) peak. Recalling that the current density is related to the net reaction rate (or flux), which is the product between the reactant concentration and the rate constant, the duck-shape is qualitatively explained by an exponential increase in current density towards $E_F^0$ due to the exponential dependence of the rate constant on the overpotential and a current density decrease towards a steady-state current after the peak potential due to a decrease in reactant concentration.

The answers to Q1 and Q2 are in principle encoded in the CV quantitatively on an absolute scale, and in principle with high accuracy and precision due to the potentiostat. The answer to Q1 is encoded in the peak positions in the CV, or, in other words, in the formal potential $E_F^0$ (i.e., the center between the reduction and oxidation peaks), which can straightforwardly be related to a given reaction (upon assuming some pre-knowledge of the system). The answer to Q2 is encoded in the peak current (or peak current density), or, perhaps more intuitively, in the charge (or charge density $Q$), which is the integral of the current density $J$, i.e., $Q = v^{-1} \int J$. Typically, the integral is evaluated between the starting and final potential of the working electrode or any other potential range of interest. Conceptually, the answers to Q1 and Q2 are similarly encoded in a galvanostatic electrochemistry through the peak position in differential charge and through the charge, respectively (with the charge simply being the product between current and time).

A real measured CV will not look like the computed "perfect"/ "ideal" CV in Figure 2. At least, it will include some capacity current from double layer charging, the magnitude and voltage dependence of which is a priori not known. Perhaps more importantly and in particular for complicated systems where the processes are in principle unknown and subject of the investigation, many (unknown) electrochemical (capacitive and faradaic) reactions may occur simultaneously. Additionally, irreversible chemical reactions of the reduced or oxidized species may follow the reduction or oxidation reactions, leading to reduction or absence of the respective pair peak [11].

A "complicated"/ "real" CV is shown in Figure 3; while no details are relevant for the sake of the following arguments, the experimental details and data are provided in the caption and on Zenodo (see Data Availability section and also [12]). In Figure 3, the individual reactions are semi-arbitrarily assigned and color coded accordingly. There appear to be two reduction peaks (yellow and purple). While the yellow reaction has a higher formal potential than the purple reaction, it must be expected that the purple reaction continues while the yellow reaction has started. There also appears to be some capacitive current (green). One prominent oxidation reaction appears (blue), which may be a redox pair with the magenta peak; this, however, is, in principle, speculation. Two significantly less prominent oxidation reactions also appear (red and orange), but it is not clear if they represent the oxidation of any redox pair. Taken together, it becomes clear that there are many convoluted reactions co-occurring simultaneously. This makes the CV challenging to interpret and also answering Q1 and Q2 infeasible. Put simply, electrochemistry alone is not a good information source for mechanistic understanding even though the potentiostat delivers current and voltage signal with high accuracy and precision (this statement is admittedly a bit oversimplified because other electroanalytical techniques such as impedance spectroscopy or differential capacity analysis may provide some mechanistic understanding).

Ideally, one would like to know the answers to Q1 and Q2 (and Q3) for the individual reactions at each data point along the CV, including reactions that occur only in small fractions but that are still highly relevant. For example, parasitic side reactions in batteries may account for only fractions of a percent per cycle (Coulombic inefficiency), yet they accumulate over time and ultimately determine battery lifetime [13]. Asked in other words, what are partial reactions and their respective contributions to the total measured current at each data point along the CV quantitatively on an absolute scale? Ultimately, one would like to assign each measured electron/Coulomb by the potentiostat at any given point to a specific reaction [14]. This goal can be achieved using *operando* or *in situ* X-ray (and neutron) methods because these methods provide the answers to Q1, Q2, and Q3 quantitatively on an absolute scale. In other words, one can perform "X-ray Coulomb Counting" for the partial reactions to deconvolute and thereby interpret the complicated CV. This renders X-ray methods particularly useful to understand electrochemical systems because they measure the same property as the potentiostat – namely Coulombs – but selectively for each reaction.

This is important because quantitative information on absolute scale at each point along an electrochemical measurement is the mechanistic understanding necessary for the design of novel and improved processes and materials as well as the development of new concepts.

## III. **X-ray methods and their particular usefulness to study electrochemical systems**

This section focuses on the particular usefulness and concept of X-ray methods to answers Q1, Q2, and Q3 quantitatively on an absolute scale ("X-ray Coulomb Counting"), without spending much text on their theoretical foundation and experimental details (for these details, we refer to reader to a recent article detailing *in situ* and *operando* X-ray scattering methods in electrochemistry and electrocatalysis [15]).

Fundamentally, X-rays have the right wavelength of Ångströms and energies of keVs such that X-ray scattering and spectroscopy methods are uniquely ideal to probe atomic/molecular structures and chemical bonds, respectively. In the words of Giorgio Margaritondo, "To study something, it is better to use a probe with similar magnitude (size and energy)" [16]. Given that charge transfer is at the heart of any electrochemical reaction, one is naturally interested in atomic/molecular phenomena, mechanism, and processes. Second, and this has been exploited extensively over the past decades, X-ray methods are ideally suited for *operando* (or at least *in situ*) studies [15]. In particular hard X-rays can relatively easily be penetrate the materials of electrochemical cells and thus be used to study the respective materials within the systems in their natural environment and under reaction conditions [15, 17]. For some more details on various X-ray methods used in electrochemistry (including advantages, limitations, and experimental constraints), we refer the reader to the Supporting Information and references [15, 18].

Many other methods cannot easily just "look inside" (realistic) cells. While X-ray methods can in principle probe timescales ranging from attoseconds/femtoseconds to hours and years (and length scales ranging from sub-Ångströms to meters), the typical measurement times, depending on the method, are from hundreds of milliseconds to tens of seconds. This provides sufficient time resolution to *operando* investigate electrochemical systems, where the electrochemical test and cycling protocols typically span minutes to hours.

Perhaps most importantly, at least in the context of the line of argumentation for "X-ray Coulomb Counting" in our perspective, and sometimes underappreciated, X-ray methods provide information (i.e., answers Q1, Q2, and Q3) that is quantitative on absolute scale (while this concept has been used in the past, it has not been explicitly introduced). X-ray Coulomb Counting allows us to interpret the respective electrochemical measurements in much greater detail and enables understanding of the driving forces and reaction mechanisms – and is as such a form of spectroelectrochemistry [19]. Depending on the type of electrochemical reaction and associated scientific question, a variety of X-ray methods can be used. Examples include – but are not limited to – X-ray diffraction (XRD) to study changes in the bulk of electrodes (e.g., ion intercalation), X-ray reflectivity (XRR) to study the formation of surface layers, and X-ray absorption (XAS) to study ion concentrations. All the methods have the powerful fact in common that the measured signal can be equated (!) to moles of material of interest in the X-ray beam, which can straightforwardly, sometimes upon some assumptions/postulates, be converted to Coulombs transferred into a certain reaction, as shown below. This is because in the respective equations, the proportionality factors relating the measured signal to moles of materials are known on an absolute scale. Accordingly, *operando* X-ray methods can be used for "X-ray Coulomb Counting" to disentangle quantitatively on an absolute scale how many electrons were transferred into which reactions at a given potential, which allows for full interpretation of a CV or other electrochemical measurement. The X-ray equations enabling this are the Integrated Intensity Formula, Master Formula, and Lambeert-Beer's Law, which shall be introduced and explained in the relevant context next using electroplating of Cu as a generic example.

### i. Integrated Intensity Formula for the intensity $I_{hkl}$ of an X-ray diffraction Bragg reflection with Miller indices $hkl$

In XRD, the scattering angle-dependent scattering intensity is measured (see Figure 4(A)). The foundational equation is the Integrated Intensity Formula for a powder sample from Warren's book [20]:

$$I_{hkl} = I_0 \frac{r_e^2 \lambda^3}{16\pi R} \mathrm{LP} \frac{m_{hkl} |F_{hkl}|^2}{(v_{\mathrm{unitCell}})^2} V \qquad \text{Equation 1}$$

Here, $I_0$ is the incident intensity, $r_e$ is the classical electron radius, $\lambda$ is the wavelength, $R$ is the sample-to-detector distance, LP is the Lorentz and polarization factor, $m_{hkl}$ and $|F_{hkl}|^2$ are the multiplicity and the modulus squared of the structure factor per unit cell, respectively, $v_{\mathrm{unitCell}}$ is the unit cell volume, and $V$ is the diffracting volume of the material giving rise to $I_{hkl}$. We note that X-ray absorption is not explicitly taken into account here but can be straightforwardly calculated. As intuitively expected, the $I_{hkl}$ conceptually depends on five type of factors, namely the incident intensity ($I_0$), some physical constants ($r_e^2, \pi$), some geometrical and instrumental parameters ($\lambda$, $R$, LP), the type of material diffracting ($m_{hkl}$, $F_{hkl}$, $v_{\mathrm{unitCell}}$), and the amount of material diffracting ($V$). The first three types are in principle known quantities or can be determined straightforwardly. For any given crystal structure, $m_{hkl}$, $F_{hkl}$, and $v_{\mathrm{unitCell}}$, are known, as are the angular (or scattering vector) positions at which the Bragg reflections appear. Accordingly, upon postulating a correlation between a given Bragg reflection $I_{hkl}$ and a material (e.g., via pre-knowledge, intuition, or phase analysis via Pawley-fitting, Rietveld refinement), the $F_{hkl}$ distribution along the scattering angle encodes the answer to Q1, i.e., which reaction is occurring. Underlying is the assumption or postulation of a given reaction, the product of which is the material belonging to $F_{hkl}$. Because $V$ corresponds to the amount of material, it encodes the answer to Q2, i.e., how much of this reaction is occurring. Given that Equation 1 relates intensity to $V$ via an equal sign and all other parameters known, $V$ is known quantitatively on an absolute scale, i.e., in $m^3$. Since the material and reaction is also known (postulated above), this can be easily converted to moles or electrons/Coulombs, which is what the potentiostat measures. If the WE itself undergoes changes, the same arguments hold to answer Q3.

We note that measuring in absolute intensities is sometimes not straightforward in XRD (or wide-angle X-ray scattering (WAXS) or total scattering with pair distribution function (PDF) analysis), which is why one may use a reference scattering materials of known $F_{hkl}$ and $V$ in order to enable quantitative phase fraction analysis approaches. While not discussed here explicitly, similar equations equating scattering intensity to materials' volume exist for small angle X-ray scattering (SAXS), where measuring at absolute intensities is straightforward [21]. Moreover, relative changes of the amount of a certain material with respect to some initial amount can be measured straightforwardly.

We illustrate the utilization of the Integrated Intensity Formula using the hypothetical but generic example of electroplating of Cu from aqueous solution via voltammetry

$$\mathrm{Cu^{2+}(aq)} + 2\mathrm{e^-} \rightleftharpoons \mathrm{Cu(s)} \qquad \text{Equation 2}$$

which occurs at potentials cathodic of 0.337 V versus SHE (for example on a porous electrode). Hypothetically, $F_{hkl}^{\mathrm{Cu(s)}}$ tells us, via a Bragg reflection observed at some expected diffraction angle, which reaction has occurred, i.e., formation of metallic Cu (or potentially a form of Cu oxide). From $I_{hkl}^{\mathrm{Cu(s)}}$ we can calculate the volume $V$ of metallic Cu within the beam area $A$ (unit of $A$ is $m^2$). This can be easily converted to mass of Cu per $A$ or moles of Cu per $A$. Because we know from Equation 2 that two electrons (let us define $n$ as the number of electrons transferred) are needed to discharge $Cu^{2+}$ ions into metallic Cu, moles of Cu per $A$ can be easily converted to electrons per $A$ via

$$Q^{\text{X-ray/XRD}} = \frac{V}{A}\frac{nF\rho_{\text{mass}}}{m_{\text{molar}}} \qquad \text{Equation 3}$$

with units $C/m^2$. This is equivalent to a charge density, which the potentiostat measured, hence "X-ray Coulomb Counting". One could then for example compare the potentiostat charge density to the "X-ray charge density" for metallic Cu, and any difference can then be attributed to another electrochemical process, such as capacitive double layer charging or hydrogen discharge below 0 V versus SHE. If chemical processes occur subsequent to electroplating, such as oxidation of Cu, the situation becomes a bit more complex, but the same principle applies towards the quantification of which Cu oxides forms and how much.

### ii. Master Formula for the angle-dependent specular reflected fraction, i.e., X-ray reflectivity

In XRR, the incident angle dependent reflected fraction is measured (see Figure 4(B)). Specular implies that the incident angle $\alpha$ and exit angle $\beta$ are equal. The foundational equation is the so-called Master Formula [21, 22]:

$$R(q_z) \approx R_{\text{F}}(q_z)\left|\frac{1}{\Delta\rho^e}\int_{-\infty}^{\infty} dz \frac{\partial\langle\rho^e(z)\rangle}{\partial z} e^{-iq_z z}\right|^2 \qquad \text{Equation 4}$$

here, $q_z = \frac{4\pi}{\lambda}\sin\alpha$ is the surface normal component of the scattering vector $\mathbf{q} = \mathbf{k}_{\text{r}} - \mathbf{k}_{\text{i}}$ with $k = \frac{2\pi}{\lambda}$. $R_{\text{F}}(q_z)$ is the Fresnel XRR of an ideally smooth and abrupt interface. $\rho^e(z)$ is the surface normal electron density profile (EDP), which is related to the corresponding refractive index ($n = 1 - \delta - i\beta$) profile, which is the typical choice to describe matter in the context of XRR; far away from resonances, $\delta$ is proportional to the electron density $\rho^e$ and $\beta$ is proportional to the linear absorption coefficient. $\Delta\rho^e$ is the difference between the electron density of the substrate and the media from which the X-rays enter the interface. While the approximate sign ($\approx$) in Equation 4 stems from the Born approximation (no multiple scattering, no refraction, no absorption), the equation still describes the relationship between the XRR and $\rho^e(z)$ quantitatively on an absolute scale. The full formalism [21, 22] can also be used but the Master Formula is more user-friendly and intuitive for the sake of argument (because a simple equation can be written down). For details regarding XRR data analysis, including model reliability and off-specular/diffuse scattering analysis, we refer the reader to the seminal textbooks [21-23].

A simple EDP with relevance to electrochemistry is a layer of constant electron density $\rho^e_{\text{layer}}$ with thickness $d_{\text{layer}}$ and interface roughness $\sigma_{\text{layer}}$; the underlying substrate is also assigned a roughness, $\sigma_{\text{substrate}}$. Such an EDP is schematically shown in Figure 4(B) together with the resultant XRR, which exhibits so-called Kiessig fringes [24], which phenomenologically originate in angle-dependent constructive/destructive interference from rays reflected from the top and bottom interface of the layer. The fringes' period is inversely proportional to the layer thickness and proportional to the electron density difference. The overall decay of the signal and the damping of the oscillations is related to the interface roughnesses.

XRR data is typically analyzed by constructing an EDP and varying its determining parameters until a satisfactory agreement between the computed and measured XRR is achieved in an iterative fitting process. Many models exist to construct an EDP, with the most widely used being the layer-model (sometimes also called slab-model), in which the EDP is constructed using one to several layers that are each defined by their electron density, thickness, and interface roughness. The layer-model is used because it is often physically meaningful, is easy to use, results in acceptable fits, and is easy to interpret. Perhaps most importantly, it contains quantitative information on an absolute scale. For this purpose, a critical assumption is made, that is, the assignment of a given electron density to a given chemical composition. Let us use again the example of Cu electroplating, but this time on a smooth and flat electrode (both properties are prerequisites of XRR measurements). Before discussion the example, we briefly comment on the absolute intensities in XRR. This is simply achieved by normalizing the reflected intensity to the incident intensity, without even requiring knowledge of the photon count of the incident beam. Moreover, an internal reference is always present in XRR due to the typically known $\Delta\rho^e$. This is conceptually similar to a reference sample in conventional X-ray diffraction. It must be stressed in passing that, unlike XRD, XRR also works for works for amorphous materials.

The data analysis of the hypothetical Cu electroplating on a flat substrate may reveal an EDP consisting of one or several layers. Let us for sake of argument and simplicity assume one layer, but the following argument holds separately for each potential additional layer. Via $\rho^e$ of the layer, we can assign a composition (note the assumptions introduced above). This would for example allow is to deduce whether the layer consists of metallic Cu, some kind of oxide, or a porous form of either, answering Q1. Mixed composition layers resulting in an average density may represent a significant challenge. Via $d$ of the layer, we can compute how much of this composition is present, answering Q2 (and Q3). The product between $\rho^e$ and $d$ can be used to calculate the "X-ray charge density" via

$$Q^{\text{X-ray/XRR}} = d\frac{\rho^e}{m_e}nF \qquad \text{Equation 5}$$

which is equivalent to

$$Q^{\text{X-ray/XRR}} = d\frac{\rho_{\text{mass}}}{m_{\text{molar}}}nF \qquad \text{Equation 6}$$

which is basically Faraday's law; *vice versa*, the electrochemical charge density can be used to calculate the "electrochemical thickness" via Equation 6 (as done conventionally). $m_e$ represents the number of electrons per mole, $N_\text{A}$ is Avogadro's constant, $F$ is Faraday's constant, $\rho_{\text{mass}}$ is the mass density, and $m_{\text{molar}}$ is the molar mass (assigned to a composition as postulated above). Note that these relations naturally have the units of a charge density (C/m$^2$) and the potentiostat charge density can now be compared to the "X-ray charge density" for the layers (equivalent to described above).

### iii. Lambeert-Beer's Law for the transmitted intensity in X-ray absorption

In X-ray absorption, the transmitted fraction of X-rays passing through a sample is measured (see Figure 4(C)). The foundational equation is the Lambeert-Beer's Law

$$I_T = I_0 e^{-\mu(c,E)t} \qquad \text{Equation 7}$$

Here, $I_T$ is the transmitted intensity, $I_0$ is the incident intensity, $\mu(c,E)$ is the concentration- ($c$) and X-ray energy- ($E$) dependent linear absorption coefficient, and $t$ is the sample thickness [21]. In principle, if measured using monochromatic radiation, and in spectroscopic modality (i.e., around an element's absorption edge), one can in principle obtain quantitative information on an absolute scale upon extracting the composition via the absorption fine structure (Q1) and amount via edge jump (Q2) (somewhat similar albeit conceptually different approaches exist, see e.g., [25]). However, spectroscopic modality is not always possible due to a variety of boundary conditions. This implies that a measurement is sensitive to all compounds in the beam and only relative changes of species are possible to observe. For the Cu electroplating example, this may be the change in Cu ion concentration (Q1 is thus implicit) from electrode to electrode upon cell polarization. Given sufficient spatial resolution, using knowledge about the initial salt concentration and assuming mass and volume conservation, the Cu ion concentration can be determined quantitatively on an absolute scale [26-28], thus answering Q2 in that we know how much Cu ions have passed through the solution.

Before we discuss a few examples, we wish to point out a few more aspects. The discussed X-ray methods allow access to buried structures and buried interfaces/interphases. Moreover, they can be performed in a variety of different 2D and 3D microscopy modalities and with surface sensitivity. As indicated above, they can be used in *in situ* and *operando* modality allowing measurements during the electrochemistry. We note that the precision and accuracy demanded by and possible with X-ray Coulomb Counting depends on the scientific question and needs to be evaluated together with statistical and systematic error propagation for each study; the latter includes uncertainties due to assumptions, the models used, and possible sample-related factors like amorphicity. For instance, in the example below where the objective was to quantify the fractional capacity losses during Extreme Fast Charging (XFC) of Li-ion batteries (LIBs) due to Li metal plating, loss of active anode or cathode material, or electrolyte decomposition, percent level accuracy may be sufficient. We stress that the same concepts apply for the neutron counterpart methods, subject to the well-established advantages and limitations of X-ray and neutron probes [21, 29]. X-ray Coulomb Counting is complementary to electrochemical mass spectrometry (and similar other methods) and it would be a significant experimental advancement to combine the two [30, 31]. Lastly, we note the implicit assumption that the probed region is representative of the entire electrode or cell, which may in general be non-applicable as a result of local overpotential and reaction delays (e.g., due to X-ray window effects [32]) or X-ray-induced effects such as radiation damage or X-ray-induced chemistry [33]. This can be overcome by mapping the entire electrode area (see below) or via a variety of mitigation strategies, for the details of which we refer the reader to two recent seminal review articles [15, 18]. Heterogeneity across the electrode thickness – the direction along which transmission measurements inherently average electrochemical reactions – can be addressed, for example, by 3D X-ray diffraction microscopy [34] or scanning-based cross-sectional mapping [35].

## IV. **Examples**

In this section, we provide a few examples of how XRD, XRR, and X-ray absorption were used for X-ray Coulomb Counting, focusing both on our own and others' work on batteries over the past decade. We will mostly address the relevant X-ray Coulomb Counting parts of the examples and refer the readers to the references for the specific hypotheses and science questions as well as experimental details and scientific findings and conclusions of the works. The XRD part will mainly concern the quantification of different loss channels during XFC of LIBs. The XRR part will mainly concern quantifying the nucleation, growth, and evolution of the solid electrolyte interphase (SEI) on model electrodes. The X-ray absorption part will mainly focus on quantifying ion concentration profiles to understand ion transport. Albeit conceptually somewhat different, we will briefly also discuss how X-ray photon correlation spectroscopy (XPCS) can be used to measure absolute velocities and how this was used to study ion transport.

### i. X-ray diffraction

XRD-based X-ray Coulomb Counting or similar approaches have been recently used by several investigators to quantitatively study (de)lithiation processes and degradation phenomena in LIBs, *inter alia* upon XFC [36-47]. The main scientific question evaluated was under which conditions parasitic Li plating occurred (Q1) and to quantify parasitic Li metal plating (Q2), its heterogeneity, and how this is related to capacity fading. Most studies were performed *in situ* (i.e., after XFC) in pouch cells using high energy X-rays, such as to easily penetrate the cells while enabling measuring a large scattering vector range. We review two examples here.

Paul et al. [42, 44] used synchrotron high energy XRD (HEXRD) to study the loss channels (loss of Li inventory, loss of active material, electrolyte decomposition) during XFC in single layer graphite/NMC pouch cells. Using spatially resolved HEXRD scanning microscopy, the authors measured diffraction peaks associated with Li metal and lithiated graphite (various stages) as well as cathode material. We note the inherent advantage of (HE)XRD through pouch cells in that all crystalline phases present in the cathode and anode are captured. Given the known areal NMC loading (that means known NMC volume $V$ in Equation 1), the NMC diffraction intensity was used as an internal reference to quantify the volumes of all other species, which included Li (via the Li (110) reflection) and lithiated graphite (via the graphite (002) and $LiC_6$ (001) reflection). Peak assignment and indexing answers Q1 and Q3. Using the relationship in Equation 3, the authors calculated how much charge transferred into each of the reactions, with a specific focus on how much Li plating occurred (in moles/m$^2$ or C/m$^2$, in this case spatially resolved, see Figure 5(A)), thereby answering Q2 via the XRD-based "X-ray charge density". By using the known relationship between lattice parameter and degree of lithiation in NMC, the authors were also able to assign how much Li was lost due to cathode particle cracking. Ultimately, the X-ray Coulomb Counting-derived "X-ray charge density" could be quantified on an absolute scale how much of the overall lost charge was due to parasitic Li plating and how much was lost due to cathode particle cracking. Relating this to the electrochemically measured lost charge allowed the authors to also quantify how much charge was lost to SEI reactions. Post-mortem analysis with optical microscopy was consistent with the HEXRD results on Li plating, with spatial correlations between metallic Li deposits detected via HEXRD and the bright regions indicative of plating seen in optical images (see Figure 5(A)). Ultimately, HEXRD provided quantitative data on phase concentrations and crystallographic details, enabling the precise measurement of Li plating and lithiation states. Another advantage of HEXRD is its ability to quantify material changes in Li-ion batteries (LIBs) non-destructively, eliminating the need for tearing down the cell. This feature enables *in situ* analysis of fully assembled cells, preserving the integrity of electrode structures and avoiding artifacts introduced by disassembly, and has the potential to be applied to batteries during operation (see also next example). Paul et al. used X-ray Coulomb Counting for various electrode thicknesses and charge rates [42, 44, 45] and in one case combined with complementary mass spectrometry titration [41], which led the authors to postulate that crosstalk between the electrodes leads to spatial correlations between of cathode degradation and anode SEI composition. In a similar approach, the same group quantified the relative changes in Li metal batteries to disentangle dead Li formation, Li corrosion, and electrolyte reduction on model Cu thin films electrodes [48].

Charalambous et al. used HEXRD to study the heterogeneity in electrochemical reactions during XFC [39]. The authors performed HEXRD scanning microscopy to reveal the spatial distribution of the lithiation extent across the entire electrode. For the NMC532 cathode, the (003) and (105) reflections were used to quantify the changes in lattice parameters and unit cell volume, which was a qualitative indicator of the lithiation extent (similar to Paul et al. above). On the anode side, the staged graphite reflections including $LiC_6$ (001) and $LiC_{12}$ (002) were used to identify and quantify Li concentrations in lithiated graphite, thereby answering Q1 and Q2. The spatial mapping revealed significant heterogeneity under XFC, with some regions showing higher fractions of fully lithiated $LiC_6$, while others remain dominated by intermediate phases such as $LiC_{12}$. The heterogeneity of NMC532 cathode and the graphite anode is spatially correlated, suggesting that local imbalances in charge transfer processes propagate through both electrodes. These variations are less pronounced under slow charging, indicating that charge rate is a primary driver of heterogeneity. In addition to the full electrode mapping at the end of charge or discharge, the authors conducted *operando* HEXRD measurements at three positions on the electrode during battery operation. By tracking the HEXRD peaks corresponding to staged graphite, the progression of lithiation was monitored. The relative fractions of each lithiated phase were determined using Equation 1 from the peak intensities, as shown in Figure 5(B). This study underscores the non-destructive nature of (HE)XRD and its ability to provide spatially resolved, quantitative insights into phase distributions within batteries. These capabilities make HEXRD a valuable tool for analyzing battery materials under real-world conditions.

### ii. X-ray reflectivity

Several investigators have used XRR [49-62] (and the neutron analogue neutron reflectivity [63-75]) has been used to study the interfacial electrochemistry for LIBs. While the lithiation mechanism has been studied using this approach, the main focus was on the nucleation, growth, and evolution of the SEI, the ion conducting but electron and electrolyte blocking passivation layer on LIB anodes. Due to its high sensitivity to surface / interface structures with sub-Å resolution and its ability to probe surfaces buried under electrolytes *in situ* and *operando*, XRR is ideally suited for the respective science questions regarding the composition of the layer (i.e., which reaction Q1) and how thick these layers are (Q2). For this purpose, the authors used model single crystalline and thin film electrodes in combination with specifically designed electrochemical cells (see e.g., [58]). While not all works have employed XRR-based X-ray Coulomb Counting, we discuss two examples here.

Cao et al. [56] used *operando* XRR to study the lithiation of Si electrodes, including SEI formation, the lithiation of the native oxide layer $SiO_x$, and the lithiation of crystalline Si electrode. During the experiment, the XRR data was measured at controlled potentials by galvanostatically decreasing the electrode potential to target voltages and holding the potential until the system reached steady-state. In this way, the structural changes at different lithiation potentials can be correlated with features in the CV curve, as shown in Figure 6(A), thereby answering Q1 and Q2 as well as Q3. By combining XRR with compositionally sensitive X-ray photoelectron spectroscopy (XPS), the authors identified the chemical reactions at specific potentials, such that the assignments of electrochemical reaction were not just postulation. They were thus able to pinpoint SEI structures to specific voltages and by using the X-ray Coulomb Counting approach and the X-ray charge densities were able to mechanistically interpret the electrochemical measurements, thereby also gathering insights into ion and electron transport through the SEI. This multimodal approach combining electrochemistry with *operando* XRR and *ex situ* XPS as well as first principles theory provided fundamental insights into the SEI growth mechanisms and the role of native oxide in the SEI templating and lithiation process. In related work, the same authors used XRR-based X-ray Coulomb Counting to quantify on an absolute scale how much charge is lost into SEI reactions (Q2) in Si thin film electrodes during the initial cycles [61].

The same group [57] also explored XRR-based X-ray Coulomb Counting to quantitatively correlate the surface structural evolution due to SEI formation with electrochemistry. Specifically, they investigated SEI formation on an inert SiC substrate. SiC was chosen because it does not react with Li, providing a model system to isolate SEI formation without interference from electrode lithiation reactions. For accurate and precise determination of $Q^{\text{electrochemistry}}$, the authors used a Teflon cone cell for precision electrochemical measurements. The cone cell was specifically designed so that only the surface of the working electrode and the Li counter / reference electrode were in contact with the electrolyte, eliminating parasitic currents from other components. The onset of the formation of a surface film as evidenced by XRR coincided with the reduction peak in the first cathodic scan in the cone cell. Using *operando* XRR, the authors obtained the thickness and electron density of the formed SEI layer during CV scan (Figure 6(B)). The electron density of the SEI layer matched that of LiF. Assuming that SEI was composed of LiF, the X-ray charge density $Q^{\text{X-ray/XRR}}$ consumed during SEI formation could be calculated from the film thickness (Equation 5/6), and the result of comparing $Q^{\text{X-ray/XRR}}$ with $Q^{\text{electrochemistry}}$ is shown in Figure 6(B), thereby answering Q1 and Q2. This agreement, together with XPS and grazing incidence XRD results, reinforced the conclusion that the majority of the current during SEI formation is due to LiF growth. This quantitative correlation between XRR measurements and electrochemical data enables precise determination of the contributions of specific reactions to the overall current. Ultimately, X-ray Coulomb Counting bridges structural characterization with electrochemical processes.

To date, XRR-based X-ray Coulomb Counting has only been carried out using mm-sized model electrodes. We denote this the "model system gap" (adapted from "pressure gap" in surface science [76]) and envision that it can be overcome by using nm-sized X-ray beams at 4$^{\text{th}}$ generation light sources toward investigating the surface electrochemistry of realistic nm–μm-sized electrode particles. This general approach was pioneered by Stubbs et al. for crystal truncation rods [77] and recently extended by Vonk et al. to XRR from μm-scaled surfaces using nanobeams [78].

### iii. X-ray absorption

Several investigators [27, 28, 79-82] have recently used *operando* X-ray absorption microscopy-based X-ray Coulomb Counting (both in scanning and full field mode) or related methods to study the ion concentration profiles in polarized electrochemical cells to understand ion transport in polymer-based and liquid electrolytes for LIBs via dilute and concentrated solution theory [83]. These studies typically employ specialized model Li|Li symmetrical electrochemical cells in which the electrodes have a distance of several mm. In these cases, the answer to Q1 is implicit, i.e., Li is moving through the electrolyte, but the question is how much (Q2).

Steinrück et al. [79] used scanning mode X-ray absorption microscopy to quantify the Li concentration profile from electrode to electrode upon cell polarization in a PEO/LiTFSI electrolyte with Li:EO ratio of 1:10, thereby answering Q2. The time-dependent concentration profiles were compared to Newman's concentrated solution theory using different transport parameters as input (see Figure 7(A)), with a specific focus on the transference number [84], which is notoriously difficult to measure in concentrated electrolytes [83] and literature values for the PEO/LiTFSI system diverge [85]. Combining the concentration profile results with ion velocity measurements (see next section) and molecular dynamics (MD) simulations, the authors concluded that the transference number is relatively independent of concentration and is about 0.2.

In a related example, Abdo et al. [26] used full-field high resolution X-ray absorption microscopy to quantify the ion concentration profiles in the same electrolyte at a Li:EO ratio of 1:5 between two Li-In alloy electrodes. They used these data to calculate the spatiotemporal electric potential "inside" the electrolyte of the polarized cell (see Figure 7(B)), thereby answering Q2. They revealed and quantitatively distinguished two contributions to the cell potential. One concentration overpotential that is governed by the potential across concentration cells and one an Ohmic contribution governed by ionic conductivity. Most of the time, the latter dominates, which led the authors to suggest research efforts towards designing new polymer-based electrolytes that minimize concentration gradients.

### iv. X-ray photon correlation spectroscopy

Even though it is conceptually somewhat different from the concepts introduced above, we briefly mention and exemplify that XPCS is also an X-ray method that allows quantification on absolute scale, albeit not in X-ray Coulomb Counting style. XPCS can be used to measure absolute velocities by employing a static reference scatterer in heterodyne modality. This introduces a Doppler shift between the scattering signals from the static reference and the sample moving at constant velocity [86, 87]. Steinrück et al. [79] and Galluzzo et al. [88] used this approach (heterodyne XPCS microscopy) to measure spatiotemporal absolute ion and electrolyte velocities in polarized Li|Li symmetrical cells to understand ion transport with a focus on transference number. Of note is the sensitivity to velocities down to Å/s.

## V. Conclusions and outlook

We provided our perspective how X-ray methods can be used to quantify the charged transferred into partial reactions in electrochemical systems on an absolute scale. Further, we explained why this approach – denoted "X-ray Coulomb Counting" – is particularly useful to understand electrochemical systems because it allows to mechanistically interpret the electrochemical signal in which the charge is measured with high accuracy and precision by a potentiostat but is itself not a good source of mechanistic understanding. Despite sometimes a bit underappreciated, X-ray Coulomb Counting combined with the *in situ* and *operando* capabilities of X-ray methods render X-rays a powerful tool to understand electrochemical systems. Towards this end, we showed a few X-ray Coulomb Counting examples from recent (LIBs) literature and hope that this approach will be expanded. While certainly extendable from LIBs to other batteries (e.g., redox flow batteries) and to electrochemical devices such as electrolyzers, supercapacitors, fuel cells, electrochemical water desalination reactors, and electrosynthesis reactors, we expect that X-ray Coulomb Counting is also applicable to photo-electrochemistry and thermocatalysis. While conceptually not as obvious, the non-X-ray signal does not necessarily need to be electrons/Coulombs but can in principle be any other reaction species.

## VI. Supporting Information paragraph

Supporting information: Table summarizing X-ray diffraction (XRD), X-ray reflectivity (XRR), and X-ray absorption spectroscopy (XAS) techniques across the main geometries and modalities.

## VII. Data availability

The python computer code and data to produce Figure 2 and Figure 3 are available at https://doi.org/10.5281/zenodo.18080283.

## VIII. Acknowledgement

We thank Michael F. Toney for mentoring us over the past years and Xiaokun Ge for the measurements shown in Figure 3. H.G.S. acknowledges funding by the German Federal Ministry of Research, Technology and Space (BMFTR) and the Ministry of Economic Affairs, Industry, Climate Action and Energy of the State of North Rhine-Westphalia through the project HC-H2. This project received funding from the European Union's Horizon Europe research and innovation programme under grant agreement No 101104032—OPINCHARGE; we acknowledge Battery2030+ for their support of OPINCHARGE.

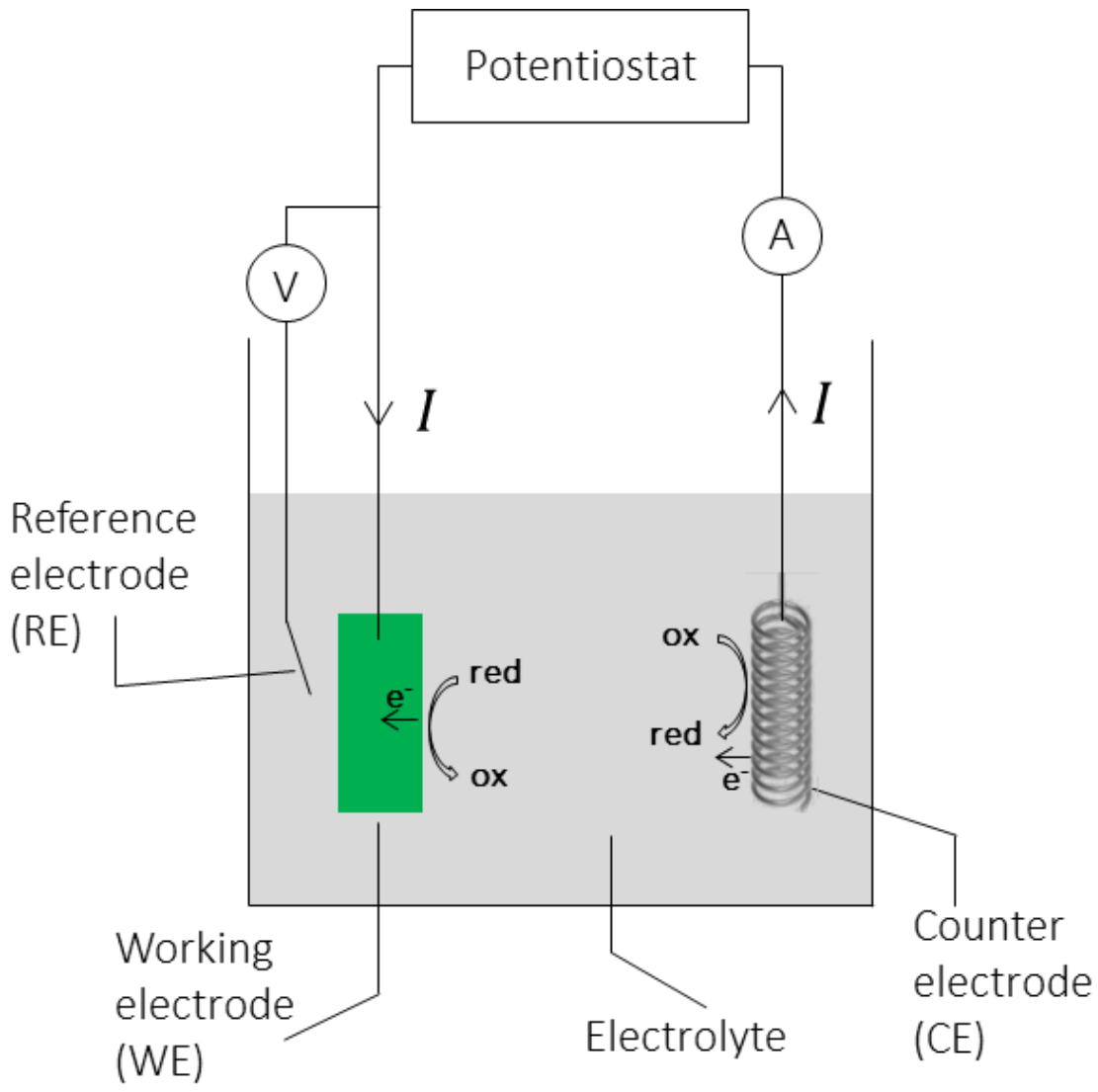

Figure 1: Generic setup for a three-electrode electrochemical cell. The voltage $E$ is controlled/measured (Ⓥ) between the working electrode (WE) and reference electrode (RE) (or counter electrode (CE) for a two-electrode cell); the current $I$ flows (Ⓐ) between the WE and CE and is sometimes given per area, weight, or volume of the WE, yielding the current density $J$.

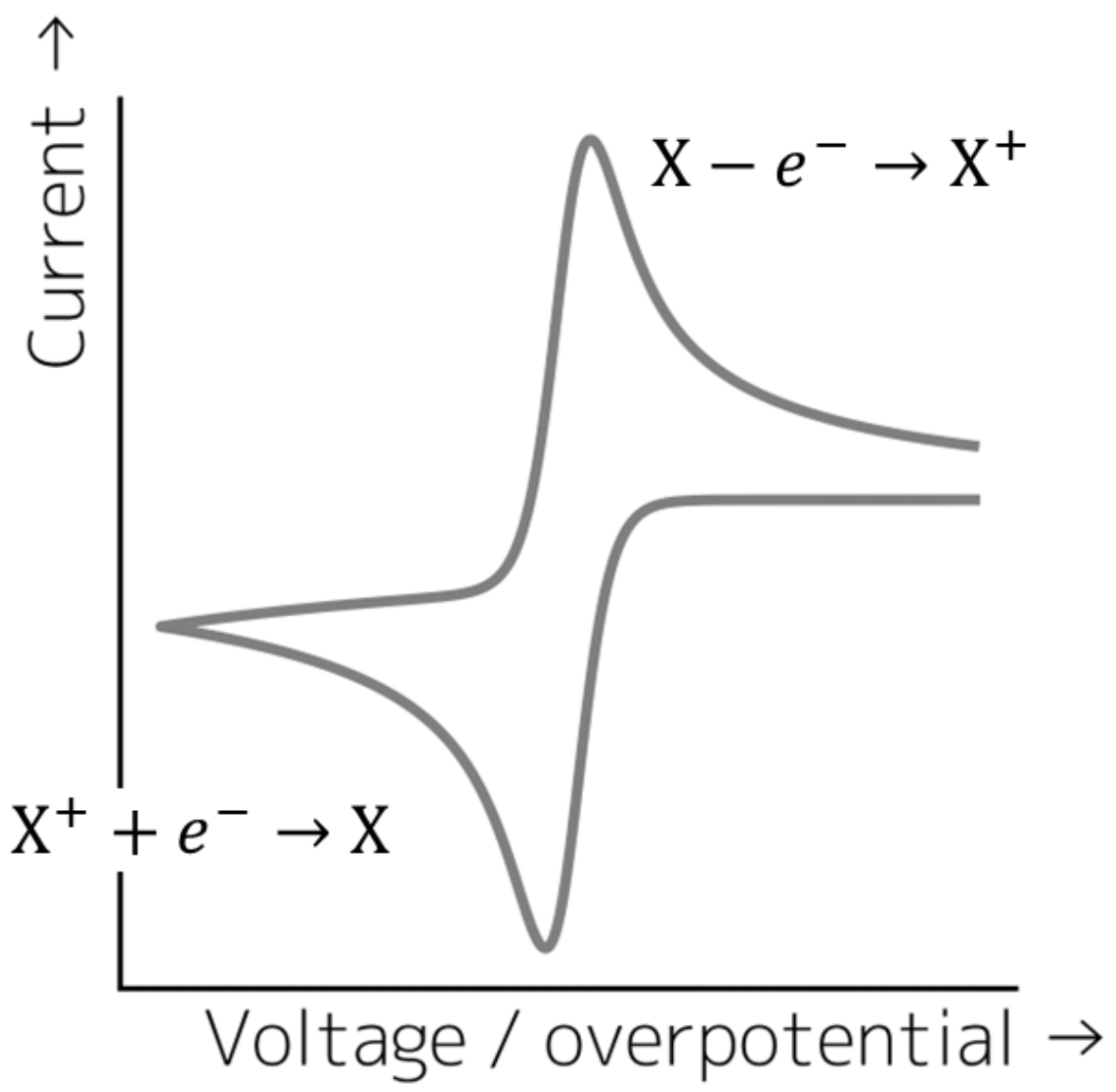

Figure 2: Typical "perfect"/ "ideal" cyclic voltammogram (CV), which was computed using Butler-Vollmer kinetics (code from [10] for macroelectrode) for a simple one-electron solution species redox reaction ($X^+ + e^- \rightleftharpoons X$). The CV exhibits the typical "duck-shape" with a cathodic (reduction) peak and an anodic (oxidation) peak.

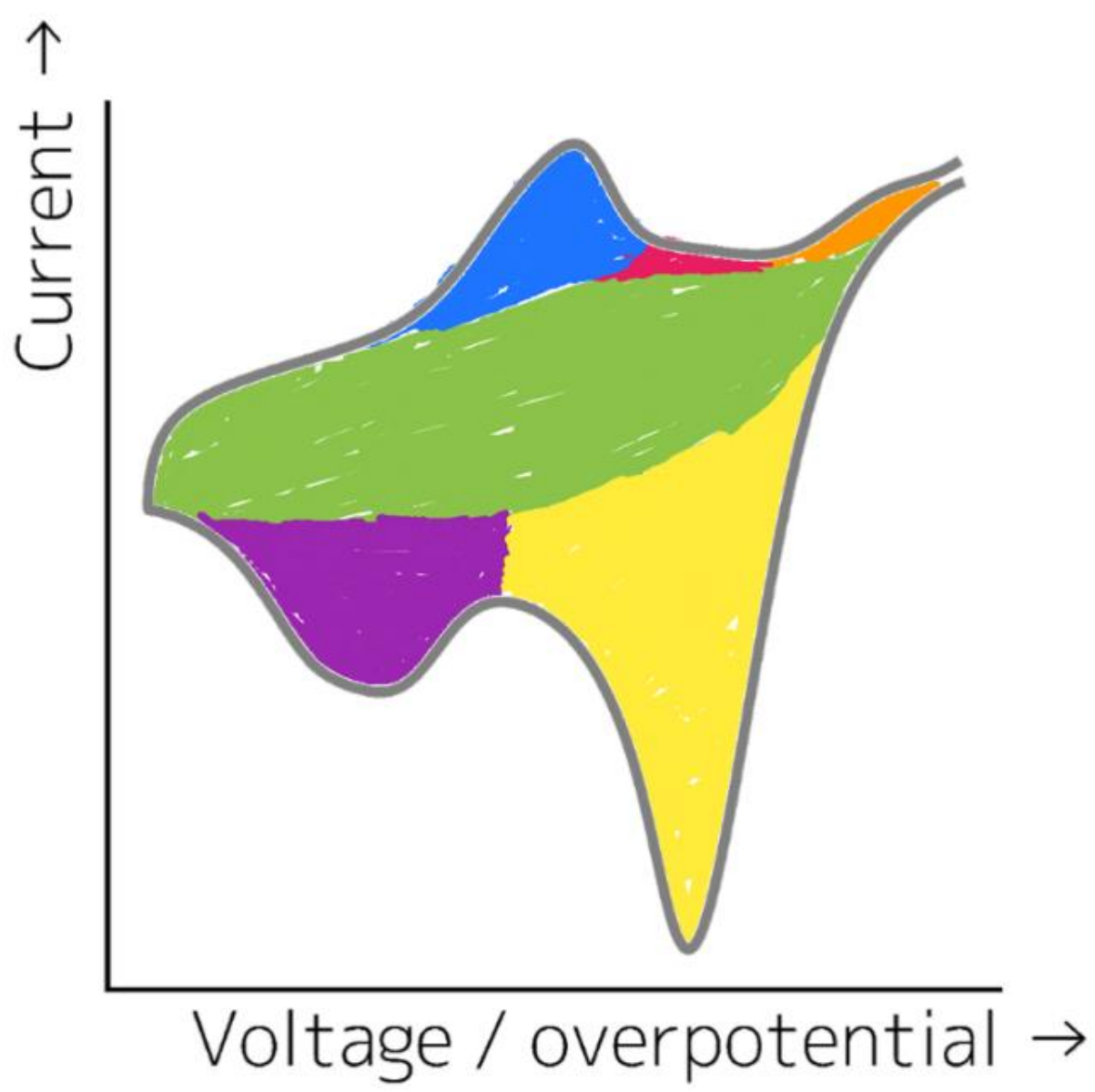

Figure 3: Typical "complicated"/ "real" cyclic voltammogram (CV), where individual reactions are semi-arbitrarily assigned and color coded. While the details of the system and experiment are irrelevant for the sake of argument and the conceptualization of X-ray Coulomb Counting, we nonetheless provide them. Working electrode: Si wafer coated with 200 nm Pt. Counter and reference electrode: Li metal. Electrolyte: LP40 (1.0 M $LiPF_6$ in ethylene carbonate (EC): diethyl carbonate (DEC) at 50:50 volume percent). Sweep rate: 50 mV/s. Voltage range: 1.0 V to 2.8 V versus $Li/Li^+$. The raw data is available on ZENODO (see Data Availability section) and further details can be found in [12].

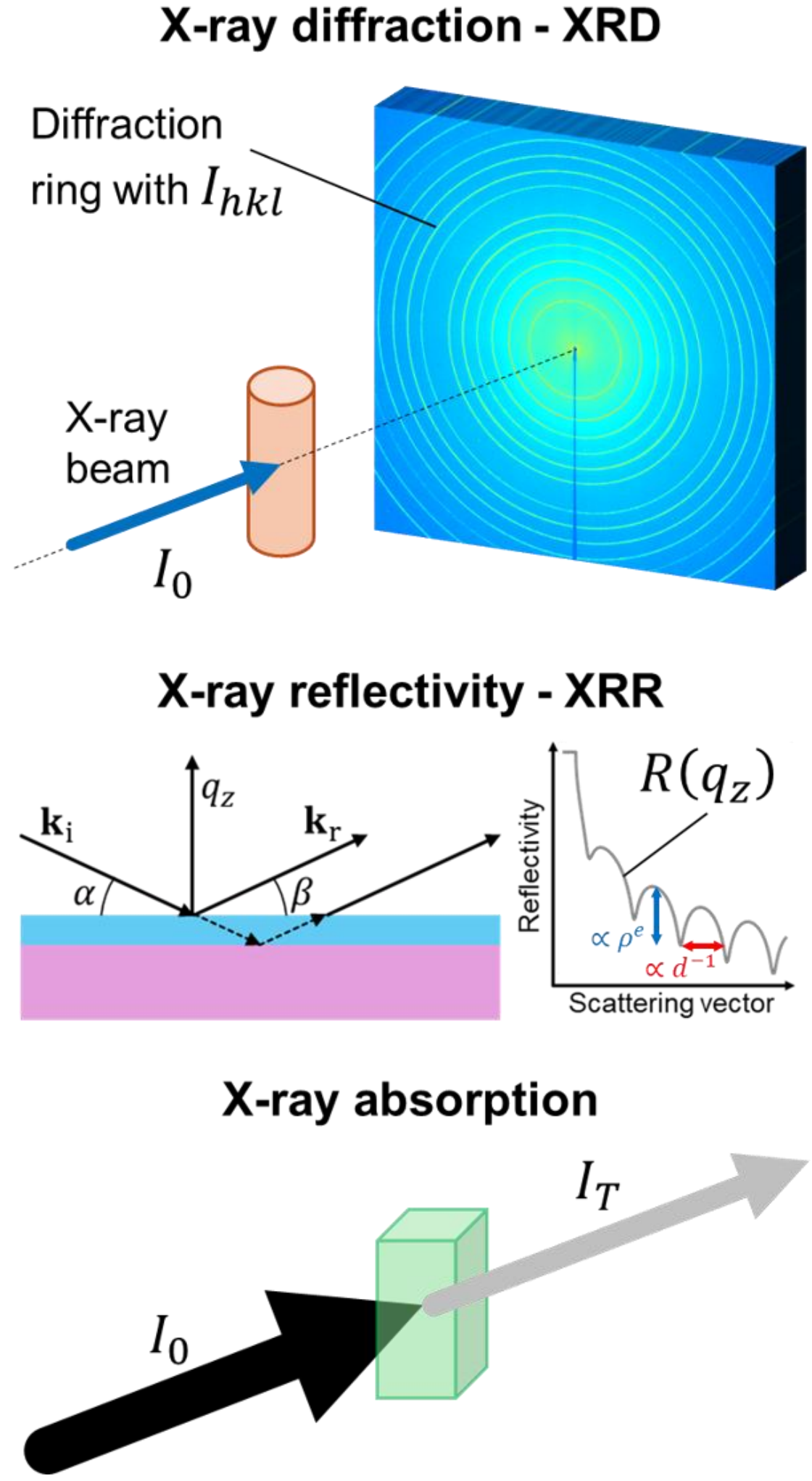

Figure 4: Schematic illustration of the three X-ray methods discussed in this work in the context of X-ray Coulomb Counting. For each of the method, the measured signal, $I_{hkl}$ for X-ray diffraction, $R(q_z)$ for X-ray reflectivity, and $I_T$ for X-ray absorption, can be related to the "X-ray charge density" that has been transferred into a given reaction (see main text for detailed explanation of the concept).

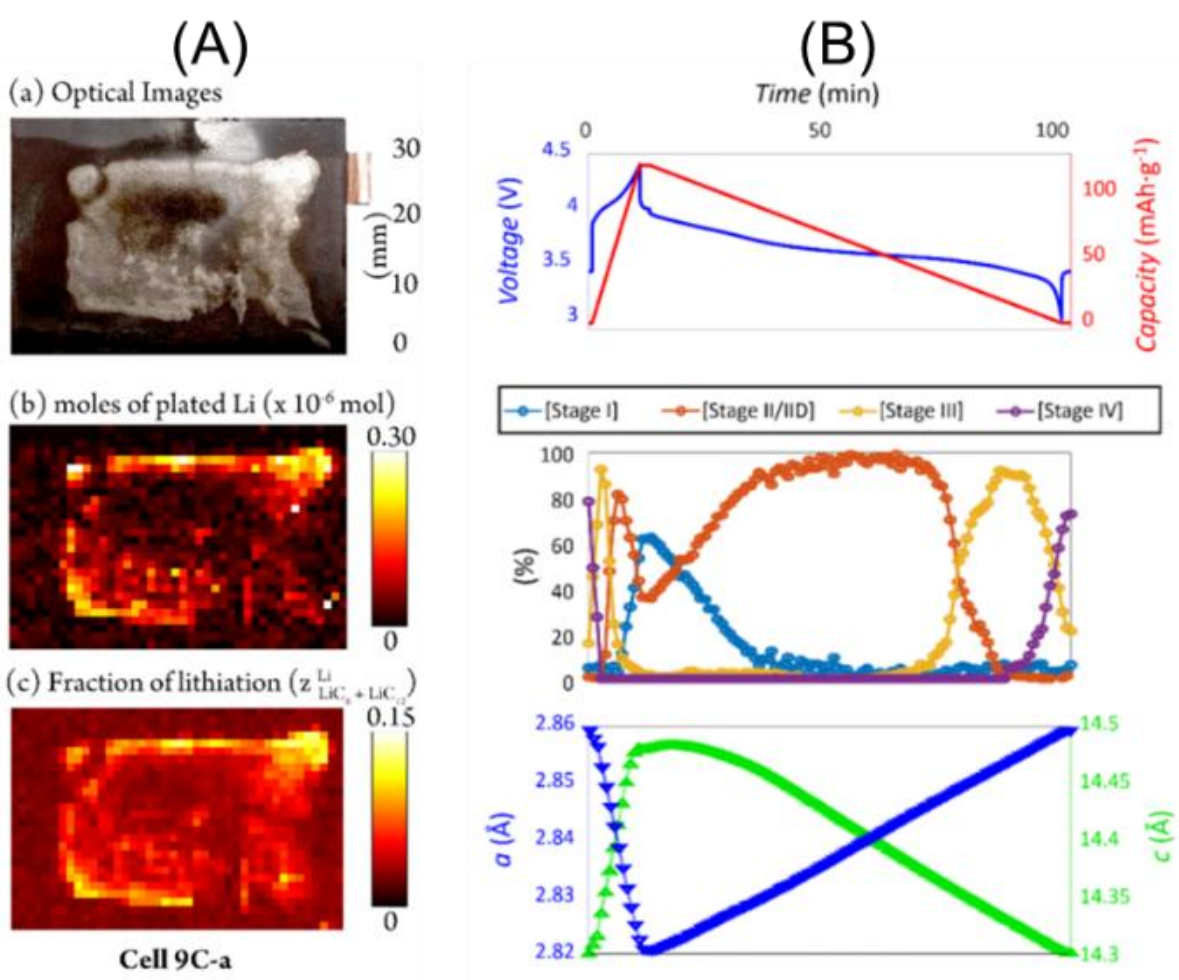

Figure 5: XRD-based X-ray Coulomb Counting. (A) Irreversible Li loss upon Extreme Fast Charging (XFC) of an NMC/graphite pouch cell, qualitatively shown via an optical image of the anode of a disassembled cell and quantitatively shown via high energy X-ray diffraction (HEXRD) microscopy (here in moles of Li metal of the anode and fraction of lithiation of graphite; Reprinted (adapted) with permission from [42]. Copyright 2021 American Chemical Society. We refer the reader to [42] for more details. (B) Quantification of anode stages and cathode lattice parameters at a single position of a cell during an XFC cycle. Reprinted from [39], Copyright (2021), with permission from Elsevier. We refer the reader to [39] for more details.

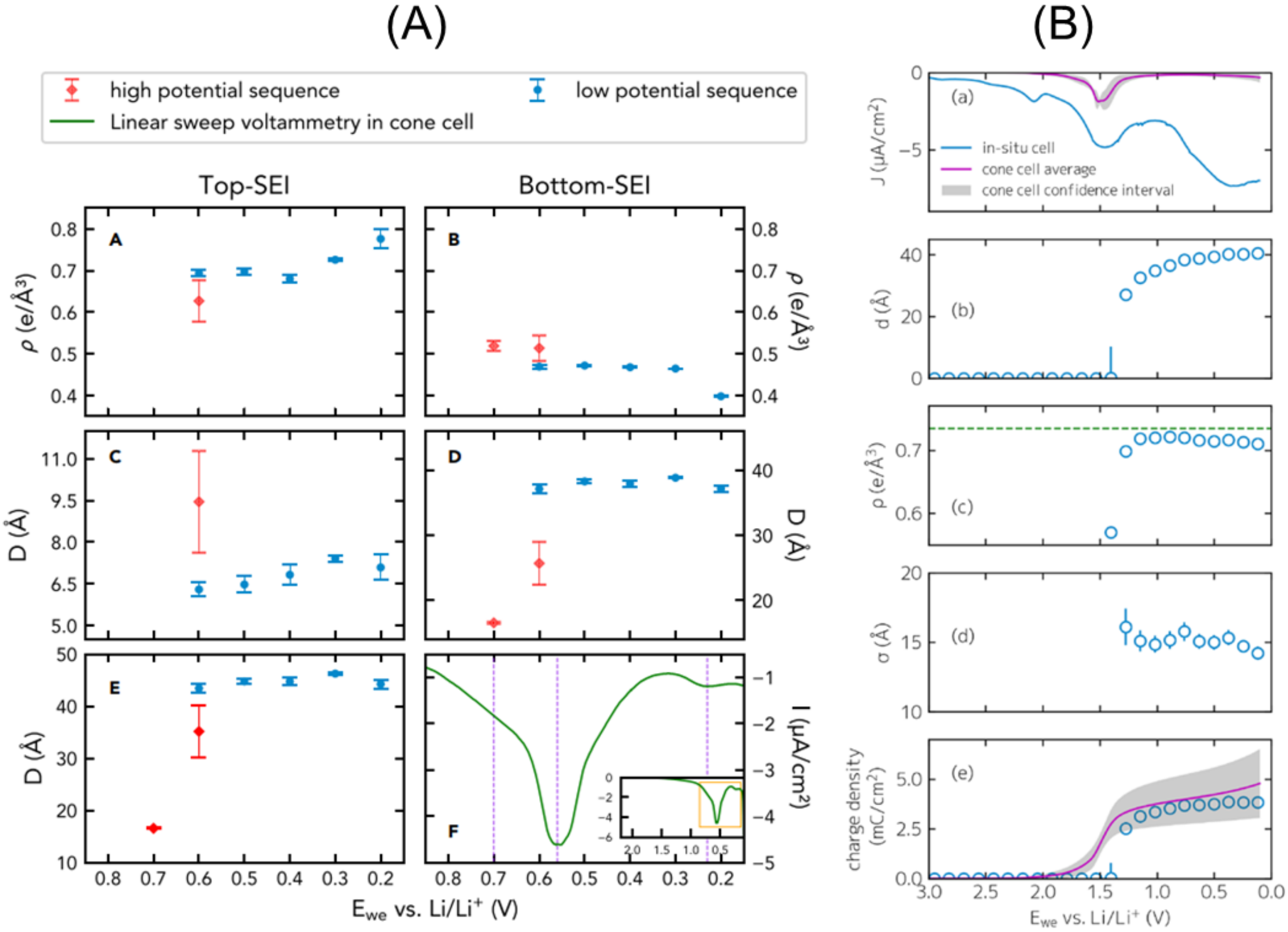

Figure 6: XRR-based X-ray Coulomb Counting. (A) Quantitative correlation of the structural changes of the Si electrode with CV profile focusing on the thickness and electron density of the solid electrolyte (SEI). Reprinted from [56], Copyright (2019), with permission from Elsevier. We refer the reader to [56] for more details. (B) Electrochemistry and SEI formation on SiC electrode. The CV and the structural properties of the SEI (thickness, electron density, and roughness are shown), and importantly a comparison between the “X-ray charge density” and the charge density measured by electrochemistry. Reprinted with permission from [57]. Copyright 2021 American Chemical Society. We refer the reader to [57] for more details.

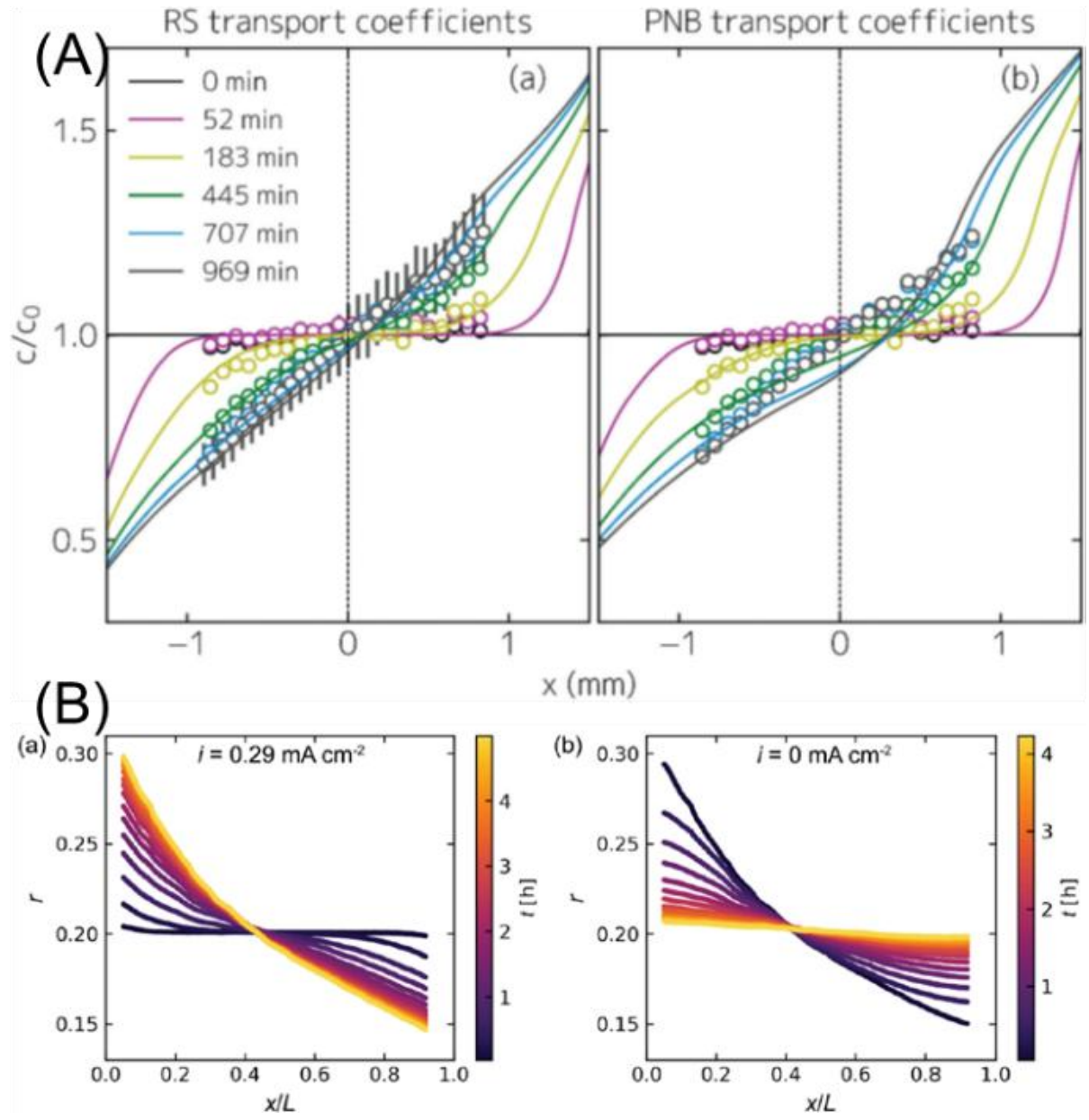

Figure 7: X-ray absorption-based X-ray Coulomb Counting. (A) Quantification on absolute scale of spatiotemporal salt concentration in a polarized symmetric cell with PEO/LiTFSI electrolyte and comparison to concentrated solution theory models using two sets of transport parameters as input. Reproduced from [79] with permission from the Royal Society of Chemistry. We refer the reader to [79] for more details. (B) Quantification on absolute scale of spatiotemporal salt concentration in a polarized symmetric cell with PEO/LiTFSI electrolyte as well as subsequent evolution during cell relaxation. Reprinted with permission from [80]. Copyright *2025* American Chemical Society. We refer the reader to [80] for more details.

# Supporting Information for

# "X-ray Coulomb Counting" to understand electrochemical systems

Chuntian Cao[1], Hans-Georg Steinrück[2,3,*]

[1]Artificial Intelligence Department, Brookhaven National Laboratory, Upton, NY 11973, USA

[2]Institute for a sustainable Hydrogen Economy (IHE-1), Forschungszentrum Jülich GmbH, An der Deutschen Welle 7a, 52428 Jülich, Germany

[3]Institute of Physical Chemistry, RWTH Aachen University, Landoltweg 2, 52074 Aachen, Germany

*corresponding author: h.steinrueck@fz-juelich.de

Table S 1 summarizes X-ray diffraction (XRD), X-ray reflectivity (XRR), and X-ray absorption spectroscopy (XAS) techniques across the main geometries and modalities, including probe size, information depth, spatial resolution, advantages and limitations, and experimental constraints. We stress that many more specialized geometries and modalities are nowadays available at 4th generation light sources — including tomographic modalities of various techniques — and hope this will help readers select the most appropriate technique for their specific electrochemical questions.

Table S 1: Overview of major X-ray techniques and associated modalities used in electrochemical studies.

| Technique | Primary Information | Geometry | Information Depth | Spatial Resolution | Advantages | Limitations | Experimental Constraints |
|---|---|---|---|---|---|---|---|
| XRD (Transmission) | Long-range crystalline order | Transmission | Bulk (µm–mm) | Beam-size dependent (mm – µm; nm possible) | Quantitative phase analysis; *operando* full-cell compatibility | Insensitive to amorphous phases | Modified cells with X-ray transparent windows (unless high energy X-rays) |
| GI-XRD | Surface sensitive structural order | Grazing incidence | nm – tens of nm | Laterally averaged (footprint-limited) | Interface-sensitive structural analysis | Requires smooth surfaces | Specialized grazing-incidence cell; precise angular alignment |
| XRR | Film thickness; electron density; interfacial roughness | Specular low-angle reflectivity | Few hundred nm | Vertical: sub-nm; laterally averaged | Sub-nm interface resolution; quantitative thin-film modeling | Requires smooth, laterally homogeneous films | Specialized reflectivity cell; unsuitable for rough composite electrodes |
| XAS (Transmission) | Element-specific oxidation state, local structure | Transmission | Bulk | Beam-size dependent | Quantitative bulk absorption; linear response | Limited sensitivity to minority species | Modified cells with X-ray transparent windows; sample thickness optimization |
| XAS (Fluorescence) | Element-specific oxidation state, local structure | Fluorescence detection | Bulk | Beam-size dependent | High sensitivity; suitable for thick samples and dilute species | Self-absorption effects | Modified cells with X-ray transparent windows |
| X-ray Microscopy | Spatially resolved composition/chemistry | Transmission or fluorescence with focused beam (or full-field) | Bulk | µm – tens of nm | Direct heterogeneity mapping | Flux-limited; long acquisition times | Requires focused beam; thin or windowed cells |